\documentclass[prd,aps,nofootinbib,floatfix,10pt]{revtex4}
\usepackage{CJK,fancyhdr}
\usepackage{amsmath,graphicx,epsfig,amssymb,dsfont,mathtools}
\usepackage[usenames]{color}
\usepackage{ulem} %% for strike-through
\usepackage{bigstrut}
\usepackage{slashed}

\begin{document}
\begin{CJK*}{UTF8}{gbsn}

\title{On the production of hidden-flavored hadronic states  at high energy}
\author{Wei Wang  }
\affiliation{INPAC, Shanghai Key Laboratory for Particle Physics and Cosmology,  MOE Key Laboratory for Particle Physics, Astrophysics and Cosmology,  \\ School of Physics and Astronomy, Shanghai JiaoTong University, Shanghai  200240,   China 
}

\begin{abstract}
I discuss the production mechanism  of hidden-flavored  hadrons at  high energy.  Using   $e^+e^-$ collisions and  light-meson pair  production in high energy exclusive processes, I demonstrate that   hidden quark pairs do not necessarily participate in short-distance  hard scattering.  Implications   are then explored in a few examples. Finally, I discuss     the  production mechanism of $X(3872)$ in hadron collisions, where some misunderstandings have arisen in the literature.  
\end{abstract}  

\maketitle

\section{Introduction} 
 
In the past few decades, hadron physics, in particular the study of exotic hadrons,  has  been the subject of extensive theoretical and experimental interest.  For recent reviews, see Refs.~\cite{Chen:2016qju,Guo:2017jvc,Ali:2017jda}.   On the experimental side, quite a number of candidates for exotic hadrons have been observed. Those include not only mesonic states like the charged $Z_c(3900)^\pm$~\cite{Ablikim:2013mio,Liu:2013dau} as four-quark candidates but also baryonic states $P_c(4380)$ and $P_c(4450)$~\cite{Aaij:2015tga}, which are likely pentaquarks. These exciting  experimental observations have stirred much   theoretical interest~\cite{Chen:2016qju,Guo:2017jvc,Ali:2017jda}. On the one hand, QCD allows  scenarios other than the usual scheme  of a meson made of  quark-anti-quark   and  a baryon as a system of three quarks.  On the other hand, due to the nonperturbative nature,  it is very difficult to have a model-independent analysis of  the internal structure of candidates for these hadrons.

The production of hadrons  at high energy  typically involves several  different scales.  It is widely believed that the hard momentum exchange   is calculable using perturbation theory.  
However,  recently there has been a debate on how to understand the  production of the $X(3872)$~\cite{Bignamini:2009sk,Albaladejo:2017blx,Esposito:2017qef}. 
The fact that   $X(3872)$   can be copiously produced in hadron collisions has led to suspicion of the molecular assignment of the $X(3872)$.   Reference~\cite{Bignamini:2009sk} has used Monte Carlo simulation  and calculated the production rates of $D\bar D^*$. Using a momentum cutoff set by the binding energy, the authors have found the simulated cross section is smaller than the data by orders of magnitude.   Such a choice of the momentum cutoff is questioned in Ref.~\cite{Albaladejo:2017blx}, while a comment on this questioning appeared  in Ref.~\cite{Esposito:2017qef}.  In addition, the authors of Ref.~\cite{Esposito:2017qef} have used  the production data of a deuteron (a loosely bound state) in a previous study~\cite{Esposito:2015fsa}. They  argued that  if  $X(3872)$  is also a  molecule,  one  expects the production of $X(3872)$ and deuterons to have  similar   behaviors.  Comparing   the data on   production of deuterons and $X(3872)$  at hadron colliders, they have found differences and thus argued   that  ``The results suggest a different production mechanism for the X(3872), making questionable any loosely bound molecule interpretation"~\cite{Esposito:2015fsa}.

In this work, I will  show that the production mechanism of the $X(3872)$ is not properly understood in Refs.~\cite{Esposito:2015fsa,Esposito:2017qef}. To do so, I will first use  standard processes, the $e^+e^-\to \rho^0\pi^0$ and $B,D$ decays, and show  that high energy production does  not  always reveal the hadron's low-energy structure.  This will induce  differences in  the  production of light nuclei like the deuteron and $X(3872)$. Finally, I will briefly comment on the  production mechanism of $X(3872)$ at hadron level and propose a new conjectured mechanism.

The rest of this paper is organized as follows. In Section~2, I  point out that naive  applications   can lead to wrong interpretations of the hadron structure~\cite{Kawamura:2013iia,Kawamura:2013wfa,Brodsky:2015wza,Chang:2015ioc,Brodsky:2016uln}.  I  give  the correct way from the viewpoint of effective field theory~\cite{Guo:2016fqg,Wang:2016vir}. In semileptonic $B$ and $D$ decays into a pair of light mesons, I will show that  hidden quarks do not participate in the scattering either. This is similar to the $B_c\to X(3872)$ transition~\cite{Wang:2015rcz}.  Section 3 concentrates on the X(3872).   A short summary is presented in the last section.

\section{Hard exclusive reactions: $e^+e^-$ collision and $B,D$ decays}

In a high energy reaction, if factorization exists, short-distance and long-distance degrees of freedom decouple.
For an exclusive process like $e^+e^-\to \rho^0\pi^0$, the constituent scaling rule is  a consequence of    perturbative QCD analysis,  which  has been derived  in a number of classic papers~\cite{Brodsky:1973kr,Brodsky:1974vy,Lepage:1980fj}. In the following, I will first present a more convenient  derivation using a modern effective field theory approach, soft-collinear effective theory (SCET). 
Using SCET,  I explicitly  demonstrate  that  the naive  constituent scaling rules must be remedied  in the case  of hidden-flavored  hadrons.

\subsection{SCET}  

SCET can be used to    study processes  involving   light  hadrons at high energy~\cite{Bauer:2000yr}.  Instead of directly studying the $s$ dependence, we introduce a dimensionless parameter $\lambda= \Lambda /\sqrt{s}$ and   and count the power dependence on $\lambda$
\begin{eqnarray}
\frac{d\sigma}{dt} \sim \frac{1}{s^2} \left(\frac{\Lambda}{\sqrt{s}}\right)^n. 
\end{eqnarray}
The scale $\Lambda$ is a low-energy scale and  may be taken as $\Lambda_{\rm QCD}$ in the case of a light quark, or $m_{c/b}$ in the case of a charm/bottom quark if involved. 
 
 At high energy, the energetic  quarks or gluons are jet-like (collinear) with the typical momentum 
\begin{eqnarray}
 p = (p^+, p^-, p_{\perp}) \propto \left( \sqrt{s}, \frac{\Lambda^2}{\sqrt s}, \Lambda\right). 
\end{eqnarray}
For an energetic quark, it is convenient to split the quark field $\psi$ into two components:
\begin{eqnarray*}
\psi=\xi+\eta,\;\;\; \xi = \frac{n\!\!\!\slash \bar n\!\!\!\slash}{4} \psi, \;\;\; \eta = \frac{\bar n\!\!\!\slash n\!\!\!\slash }{4} \psi,
\end{eqnarray*}
where $n$ and $\bar n$ are two light-like vectors: $n^2=\bar n^2=0$. 
The quark field   scaling can be obtained by considering  the two-point correlator:
\begin{eqnarray}
 \langle 0|T[\psi(x)\overline \psi(y)] |0\rangle = \int \frac{d^4 p}{(2\pi)^4} e^{-ip\cdot (x-y)} \frac{i(p\!\!\!\slash+m)}{p^2-m^2+i\epsilon}. 
\end{eqnarray}
This gives  
\begin{eqnarray}
 \xi \propto \lambda, \;\;\; \eta \propto\lambda^2.
 \end{eqnarray}
For a collinear  photon/gluon field, one has the propagator in the  general $R_\xi$ gauge as:
\begin{eqnarray}
 \langle 0| T[A_\mu(x)A_\nu(y)]|0\rangle &=& \int \frac{d^4 p}{(2\pi)^4} e^{-ip\cdot (x-y)} \frac{-i }{p^2-m^2+i\epsilon} \left[g_{\mu\nu} -(1-\xi) \frac{p_\mu p_\nu}{p^2} \right]. 
\end{eqnarray}
Then one  finds the scaling: 
\begin{eqnarray}
 n_+ A\propto 1, \;\;\; A_{\perp}\propto \lambda, \;\;\; n_-A\propto \lambda^2. 
\end{eqnarray}
In the following, we will not encounter a soft gluon/photon.

A mesonic/leptonic  state scales as  $
|M\rangle  \propto \lambda^{-1}$, which can be easily derived from the normalization of states:
\begin{eqnarray}
\langle M(p)|M(p')\rangle =(2\pi)^3 2E_p \delta^3(\vec p-\vec p'). 
\end{eqnarray}
For a lepton, scalings of state and field will cancel, and thus one  only needs to consider the final hadron.

\subsection{$e^+e^-\to \rho^+\pi^-$ and $e^+e^-\to \rho^0\pi^0$}

%%%%%%%%%%%%%%%%%%%
%%%%%%%%%%%%%%%%%%%
\begin{figure}  
\includegraphics[width=10cm]{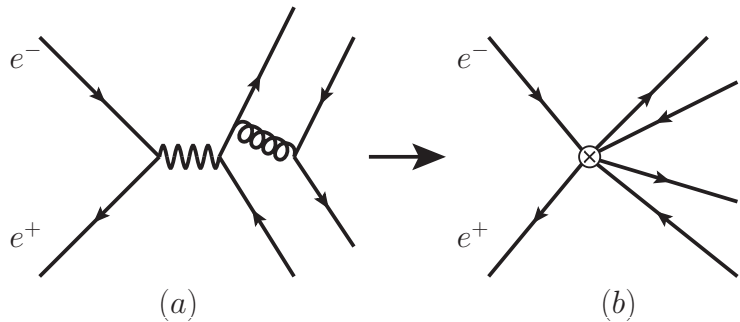}
\caption{Leading power Feynman diagram for the photon contribution to $e^+e^-\to \rho^+\pi^-$ in the full theory (a) and effective field theory (b).    }
\label{fig:EFT_VP}        
\end{figure}
%%%%%%%%%%%%%%%%%
%%%%%%%%%%%%%%%%%%%

At  high energy with $\sqrt s\gg \Lambda_{\rm QCD}$, exclusive processes are calculable   in  perturbation theory. 
When factorization holds, one may separate the interactions
according to the scales involved   using the operator product expansion.  The interactions above the factorization scale  can be  integrated out, which results in an effective field theory.     We show the matching for the $e^+e^-\to \rho^+\pi^-$ in Fig.~\ref{fig:EFT_VP}.  The photon propagator, quark propagator and gluon propagator are highly off-shell, and thus these propagators can be shrunk   to the same space-coordinate. Then in   low energy effective field theory  the cross section is factorized as:
\begin{eqnarray}
{\cal M}(e^+e^-\to \rho^+\pi^-)= C \otimes \langle\pi^+| \bar\xi_{ n}\frac{\bar n\!\!\!\slash}{2} \xi_{ n}|0\rangle\otimes \langle \rho^+| \bar\eta_{\bar n}\gamma_\perp \xi_{\bar n}|0\rangle. 
\end{eqnarray}
Since  the $\rho^+$ is transversely polarized,  the small component $\eta$ contributes. $n$ and $\bar n$ are two unit light-like vectors with $n^2=\bar n^2=0$, $n\cdot \bar n=1$.  Here $\otimes$ denotes a convolution in the space coordinate, and $C$ is an ${\cal O}(1)$  coefficient. 
Using the building blocks given in the last subsection, we have the power counting:
\begin{eqnarray}
{\cal M}(e^+e^-\to \rho^+\pi^-)\propto \lambda^3. 
\end{eqnarray} 
The cross section scales as: 
\begin{eqnarray}
 \sigma(e^+e^-\to \rho^+\pi^-) \propto \frac{1}{s} \lambda^6  \propto  \frac{\Lambda^6}{s^4}.  
\end{eqnarray}
This result is consistent with the perturbative QCD calculations~\cite{Lu:2007hr,Braguta:2008hs}, and validated by experimental data~\cite{Adam:2004pr,Ablikim:2004kv,Aubert:2006cy,Belous:2009yp,Shen:2013okm}. 
The above result  is also consistent with the classical  constituent scaling rule~\cite{Brodsky:1973kr,Brodsky:1974vy,Lepage:1980fj}:
\begin{eqnarray}
 \sigma(e^+e^-\to \rho^+\pi^-) \propto \frac{1}{s^{n_t-3}} \times \frac{1}{s}, \label{eq:consitutentscaling}
\end{eqnarray}
where $n_t$ denotes the  total number of constituents in the process. Since $\rho$ and $\pi$ contains two quarks, we have $n_t= 1+1+2+2=6$. The last factor $1/s$ arises from helicity suppression.

%%%%%%%%%%%%%%%%%%%
%%%%%%%%%%%%%%%%%%%
\begin{figure}  
\includegraphics[width=10cm]{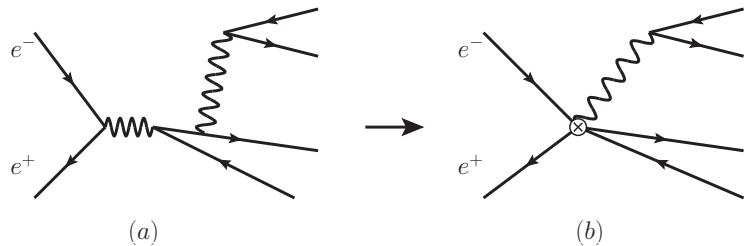}
\caption{Leading power Feynman diagram for the photon contribution to $e^+e^-\to \rho^0\pi^0$ in the full theory (a) and effective field theory (b). Unlike the  $e^+e^-\to \rho^+\pi^-$ case,  the two quarks in $\rho^0$ do not participate in the hard-scattering, and thus the  leading power amplitude is not sensitive to the two-quark nature of $\rho^0$! }
\label{fig:EFT_photon}        
\end{figure}
%%%%%%%%%%%%%%%%%
%%%%%%%%%%%%%%%%%%%

Now we consider the $e^+e^-\to \rho^0\pi^0$.  
A vector meson such as $\rho^0$  can  be produced by a photon field $
 \langle \rho^0| A^\mu_\perp|0\rangle$.  Then the   decay amplitude has the power scaling 
 \begin{eqnarray}
{\cal M} \propto  \langle \pi|\bar \xi_n \bar n\!\!\!\slash \xi_n|0\rangle \times  \langle \rho^0| A^\mu_\perp|0\rangle \propto \lambda, 
\end{eqnarray}
which leads to the  cross section 
\begin{eqnarray}
 \sigma(e^+e^-\to \rho^0\pi^0) \propto \frac{1}{s} \lambda ^2  \propto  \frac{\Lambda^2}{s^2}.  
\end{eqnarray}
This result contradicts the naive constituent scaling rule given in Eq.~\eqref{eq:consitutentscaling}.

A few remarks are in order. 
\begin{itemize}
\item 
It is necessary   to  stress that  the photon  contribution is suppressed by
the  fine structure constant $\alpha_{\rm em}$ and is less
important at low energy. 
At very high energy the photon contribution is at leading power~\cite{Lu:2007hr,Guo:2016fqg}. It has also been shown that  this mechanism will lead to important consequences in electroweak penguin-dominated $B$ decays~\cite{Beneke:2005we,Lu:2006nza}.  

\item 
To understand the above behaviors, one can  
count the valence degrees of freedom of the neutral vector meson as $n_i=1$, which amounts to counting the number of lines (a photon in this case, as shown in Fig.~\ref{fig:EFT_photon})
attached to the effective vertex shown in Fig.~\ref{fig:EFT_photon}.   
\item From   the viewpoint of effective field theory, the nonzero matrix element $ \langle \rho^0| A^\mu_\perp|0\rangle $ uses Heisenberg operators. When converting to the interaction picture, one must include the interactions: 
\begin{eqnarray}
\langle \rho^0| A^\mu_\perp|0\rangle \equiv \langle  T [\rho^0| A^\mu_\perp \times {\rm exp}\left[i\int d^4x{\cal L}(x) \right]|0\rangle, 
\end{eqnarray}
with the standard  QED interaction Lagrangian
\begin{eqnarray}
{\cal L} = \bar q  e e_q A\!\!\!\slash q. 
\end{eqnarray}

\item The scale dependence of parton distribution function (PDF) is encoded in the DGLAP evolution. For a flavor singlet, the quark PDF and gluon PDF mix with each other. From this viewpoint, our results   would be similar: the photon field  operator   at a high scale evolves to  the quark-anti-quark field operator at a low scale.  These operators will mix in the scale evolution. 

\item The above example indicates that not all constituents in the hadron participate in the hard scattering. Therefore,  one cannot use the scaling behavior of the cross section to decipher the hadron's structure. 

\end{itemize}

\subsection{ $e^+e^-\to Z_c^\pm\pi^\mp$, $e^+e^-\to D_{s0}(2317)^\pm D_s^{*\mp} $ and $e^+e^-\to\phi f_0(980)$}

After  the discussion with an ordinary hadron, I now propose a few processes  to explore the  $Z_c(3900)^\pm$~\cite{Ablikim:2013mio,Liu:2013dau}, $D_{s0}(2317)$~\cite{Aubert:2003fg}, and $f_0(980)$. 

\begin{figure}  
\includegraphics[width=10cm]{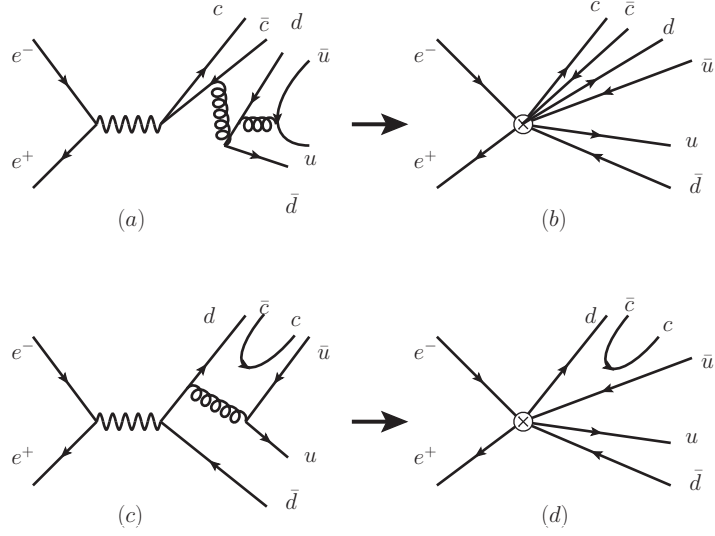}
\caption{Feynman diagrams for  $e^+e^-\to Z_c^\pm \pi^\mp $ in the full theory (a, c) and effective field theory (b, d).  }
\label{fig:feynman_Zc}        
\end{figure}

The  $Z_c(3900)^\pm$ decays into $J/\psi\pi^\pm$, and the lowest Fock state is expanded as four quarks~\cite{Ablikim:2013mio,Liu:2013dau}. If the quarks are democratically distributed, this is identified as a tetraquark meson. It is also likely that the $Z_c(3900)^\pm$ is made of two mesons, i.e. that it is a hadron molecule. 
For  $e^+e^-\to Z_c^\pm\pi^\mp$ production,  typical  Feynman diagrams are shown  in Fig.~\ref{fig:feynman_Zc}.  These two diagrams (panels (a, c)) will compete. At low energy, the two charm quarks will be produced first, and then the light quarks are generated. The perturbative suppression  for the production of light quarks might not be severe. So it is likely that the panel (a) dominates near threshold. However, at very high energy, panel (a) is suppressed due to the hard gluons, and  the leading power matrix element is given in panel (c) with the factorization formula: 
\begin{eqnarray}
 {\cal M}(e^+e^-\to Z_c^\pm\pi^\mp) = C\otimes \langle\pi^+| \bar\xi_{ n, d}\frac{\bar n\!\!\!\slash}{2} \xi_{ n, u}|0\rangle\otimes   \langle Z_c^-| \bar\xi_{\bar  n, u}\frac{  n\!\!\!\slash}{2} \xi_{\bar n, d}|0\rangle.   \label{eq:factorization_ee_Zcpi}
\end{eqnarray} 
Thus we can predict that the cross section scales as 
\begin{eqnarray}
 \sigma(e^+e^-\to Z_c^\pm\pi^\mp) \propto \frac{1}{s} \lambda ^2  \propto  \frac{\Lambda^4}{s^3}.  \label{eq:eetoZpi}
\end{eqnarray}

%%%%%%%%%%%%%%%%%%%
%%%%%%%%%%%%%%%%%%%
%%%%%%%%%%%%%%%%%
%%%%%%%%%%%%%%%%%%

\begin{itemize}
\item 
The above matrix elements in Eq.~\eqref{eq:factorization_ee_Zcpi} are written in the Heisenberg picture. When converting to the interaction picture, one has to include the interaction  below the scale $\sqrt{s}$, and   formally have 
\begin{eqnarray}
\langle Z_c^-| \bar\xi_{\bar  n, u}\frac{  n\!\!\!\slash}{2} \xi_{\bar n, d}|0\rangle   = \langle Z_c^-| \bar\xi_{\bar  n, u}\frac{  n\!\!\!\slash}{2} \xi_{\bar n, d}   \times  {\rm exp} [i\int d^4x {\cal L}_{\rm int}(x)]|0\rangle, \label{eq:expansion_Zcpi}
\end{eqnarray}
with the interaction Lagrangian:
\begin{eqnarray}
{\cal L}_{\rm int} = \bar c gA\!\!\!\slash c + \bar q  gA\!\!\!\slash q. 
\end{eqnarray}
Notice that unlike the $\rho^0$, one cannot handle  this time-ordered product perturbatively.  So Eq.~\eqref{eq:expansion_Zcpi} is a formal equation. 

\item 
The production of $\bar cc$ in panel (c) is suppressed by $1/m_{c}^2$.  Thus at low energy $\sqrt s\sim m_c$, both panel (a) and panel (c) contribute.

\item  The reaction $e^+e^-\to D_{s0}(2317)^\pm D_s^{*\mp} $  was observed for the first time with a data sample of 567~pb$^{-1}$ collected with the BESIII detector operating at the BEPC-II collider at $\sqrt s=4.6 $ GeV~\cite{Ablikim:2017rrr}. The low collision energy does not guarantee  the use of perturbation theory.  However, we expect a study at Belle-II can  uniquely  test the same scaling behavior given in Eq.~\eqref{eq:eetoZpi}.

%%%%%%%%%%%%%%%%%%%
%%%%%%%%%%%%%%%%%%%
\begin{figure}  
\includegraphics[width=10cm]{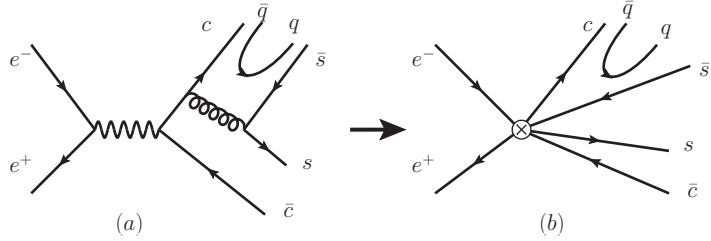}
\caption{At high energy, Feynman diagrams for the   $e^+e^-\to D_{s0}(2317)^+D_s^{*-} $ in the full theory (a) and effective field theory (b).  }
\label{fig:feynman_Ds2317}        
\end{figure}
%%%%%%%%%%%%%%%%%
%%%%%%%%%%%%%%%%%%

\item  
 $e^+e^-\to\phi f_0(980)$ can   proceed similarly, with the Feynman diagrams given in Fig.~\ref{fig:feynman_phif0}. Experimentally, the BESIII and  Babar  collaborations have used the initial state radiation and measured  $e^+e^-\to \phi \pi^+\pi^-$~\cite{Shen:2009zze,Lees:2011zi}.  I suggest that our experimental colleagues study the collision energy dependence and validate the production mechanism in this work.

%%%%%%%%%%%%%%%%%%%
%%%%%%%%%%%%%%%%%%%
\begin{figure}  
\includegraphics[width=10cm]{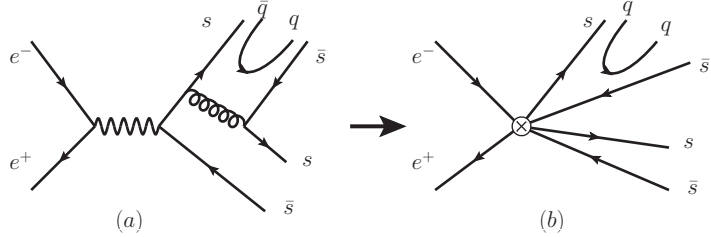}
\caption{At high energy, Feynman diagrams for   $e^+e^-\to \phi f_0(980) $ in the full theory (a) and effective field theory (b).  }
\label{fig:feynman_phif0}        
\end{figure}
%%%%%%%%%%%%%%%%%
%%%%%%%%%%%%%%%%%%

\end{itemize}

\subsection{$B,D$ decays into a light-meson pair and $\gamma\gamma$ fusion }

%\subsection{$B$ and $D$ decays} 

%%%%%%%%%%%%%%%%%%%
%%%%%%%%%%%%%%%%%%%
\begin{figure}  
\includegraphics[width=5cm]{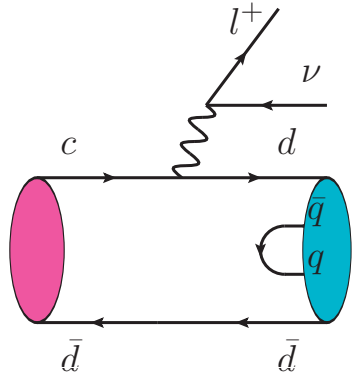}
\caption{A typical Feynman diagram for $D\to \pi^+\pi^-e^+\nu$. The leptonic sector can be calculated using perturbation theory, while the $D\to \pi\pi$ transition is parameterized in terms of   form factors.  }
\label{fig:feynman_BD}        
\end{figure}
%%%%%%%%%%%%%%%%%
%%%%%%%%%%%%%%%%%%

Pairs of  light  pseudo-scalar mesons   have  a special relation with light scalar mesons, for instance $f_0(980)$ and $\kappa(800)$~\cite{Patrignani:2016xqp}. Recently, there have been studies of  semileptonic $B, D$ decays into light meson pairs~\cite{Aaij:2014lba,Ablikim:2015mjo}. A typical Feynman diagram for $D\to \pi^+\pi^-e^+\nu$ is given in Fig.~\ref{fig:feynman_BD}, and for other channels, the Feynman diagrams are similar.  In these  decays, the leptonic sector can be factorized out  and calculated perturbatively. The nonleptonic matrix element  is then parameterized as: 
\begin{eqnarray}
    \langle (\pi^+\pi^-)_S(p_{\pi\pi})|\bar s \gamma_\mu\gamma_5 c| D  (p_D)
 \rangle  &=& -i  \frac{1}{m_{\pi\pi}} \bigg\{ \bigg[P_{\mu}
 -\frac{m_{D }^2-m_{\pi\pi}^2}{q^2} q_\mu \bigg] {\cal F}_{1}^{D\to \pi\pi}(m_{\pi\pi}^2, q^2) \nonumber\\
 &&
 +\frac{m_{D }^2-m_{\pi\pi}^2}{q^2} q_\mu  {\cal F}_{0}^{D\to \pi\pi}(m_{\pi\pi}^2, q^2)  \bigg\},
 \label{eq:generalized_form_factors}
\end{eqnarray}
where we have only shown the S-wave $\pi\pi$ final state. $m_{\pi\pi}$ is the $\pi\pi$ invariant mass.  
This 
defines the S-wave generalized form factors ${\cal F}_i$~\cite{Meissner:2013hya}. 
Here,  $P=p_D+ p_{\pi\pi}$ and $q=p_D- p_{\pi\pi}$.

The study of generalized form factors  requires knowledge of the  generalized light-cone distribution amplitude (LCDA). The leading twist LCDA of    $\pi\pi$ systems is defined by  two  quark fields. The leading-power behavior in $1/s$ is then determined by the two quarks, irrespective of the structures of the  and $\pi\pi$ systems~\cite{Diehl:2003ny}. Here 
notice the different dependence on the invariant mass  $m_{\pi\pi}$ and the collision energy $\sqrt{s}$.

Using the light-cone sum rules, one can derive the factorization formula~\cite{Meissner:2013hya}: 
\begin{eqnarray}
 {\cal F}_{i}(q^2, m_{\pi\pi}^2) =  B_0 m_{\pi\pi} F_{\pi\pi}(m_{\pi\pi}^2)  \overline   F_i (m_{\pi\pi}^2,q^2),
\end{eqnarray}
where $B_0$ is the QCD condensate parameter and $F_{\pi\pi}(m_{\pi\pi}^2) $ is the $\pi\pi$ scalar form factor. The  $ \overline   F_i (m_{\pi\pi}^2,q^2)$ is a function of the two-meson LCDA defined by
\begin{eqnarray}
 \langle \pi^+\pi^-|\bar s(x) (1, \gamma^\mu, \sigma^{\mu\nu}) s|0\rangle. 
\end{eqnarray}
This approach  has recently been  used  to calculate  heavy meson decays in Refs.~\cite{Chen:2002th,Doring:2013wka,Meissner:2013hya,Meissner:2013pba,Wang:2014ira,Wang:2015paa,Wang:2015uea,Shi:2015kha,Hambrock:2015aor} and agreements with relevant data~\cite{Aaij:2014lba,Ablikim:2015mjo} are found.

\begin{figure}  
\includegraphics[width=5cm]{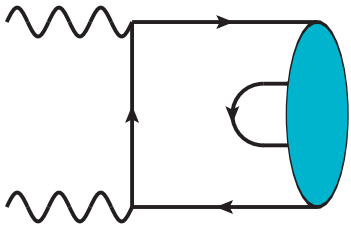}
\caption{Feynman diagrams for $\gamma\gamma\to \pi^+\pi^-/K^+K^-/\pi^0\pi^0$.  }
\label{fig:feynman_2gamma}        
\end{figure}

It should also be viable to study  two-meson production in $\gamma\gamma$ processes at BESIII~\cite{Redmer:2018gah} and Belle-II in future. Such processes are only sensitive to the leading twist generalized LCDA, and the Feynman diagram is given in Fig.~\ref{fig:feynman_2gamma}. Actually,   the Belle collaboration has published the first investigation of momentum dependence   in the two-pion system~\cite{Masuda:2015yoh}. The $\pi^+\pi^-$ system was studied for momentum transfers between $3\leq Q^2[{\rm GeV}^2] \leq 30$.  Again we should warn that  the  existing     proposals  to use  this process and  extract the structure of scalar mesons~\cite{Kumano:2017lhr,Kumano:2018jgo}   are problematic.  

%%%%%%%%%%%%%%%%%%%
%%%%%%%%%%%%%%%%%%%
%%%%%%%%%%%%%%%%%
%%%%%%%%%%%%%%%%%%

\section{$X(3872)$} 

\subsection{ Differences between production of light nuclei  and $X(3872)$} 

%A lesson learned from the above examples  is:  {\it for the high energy production  of hidden flavored hadrons,    the cross section  do not necessarily scale according to  the hadron's internal structure.} On the other hand,   {\it the contributing  matrix element  $
% \langle \rho^0| A^\mu_\perp|0\rangle$  can not lead to  the assignment  of  $\rho^0$ as a photon! }

The authors of Ref.~\cite{Esposito:2017qef} have used  production data for a deuteron (a loosely bound state) in a previous study~\cite{Esposito:2015fsa}. They  argued that  if  $X(3872)$  is also a  molecule,   a similar scaling behavior in the production rate to that for a deuteron is expected.  Using  data for the  production of deuterons and $X(3872)$  at hadron colliders, they have found differences and thus argued   that  ``The results suggest a different production mechanism for the X(3872), making questionable any loosely bound molecule interpretation"~\cite{Esposito:2015fsa}.

Actually,  there are dramatic  differences  between the production of the deuteron and the $X(3872)$.  Unlike the  deuteron, which contains  6 quarks,  the $X(3872)$ contains  hidden flavors, and thus one cannot use the same power scalings for the two hadrons. Instead,  the  production rates of  the $X(3872)$ at high energy hadron colliders are determined by the quark-anti-quark field. The production rates  do not scale according to its low energy structure, whether molecule or tetraquark.

The production of the $X(3872)$ meson involves many length scales~\cite{Braaten:2004jg}. The creation of the $\bar cc$ pair with a small relative momentum requires a hard-scattering process at  the  scale $m_c$.  This $\bar cc$ pair can be color singlet or color-octet. The evolution of the $\bar cc$ into a color-singlet hadron occurs over a softer scale $m_cv$ or $m_cv^2$. Then the evolution of the charmed mesons   occurs over an even lower scale $m_\pi$. At last the binding of $DD^*$ into the molecular state $X$ occurs over a very long length scale. To calculate the production rates of $X(3872)$ at high energy, one can  employ the  nonrelativistic QCD  (NRQCD) approach: 
\begin{eqnarray}
\sigma (p\bar p\to X )&=& \hat  \sigma(p\bar p\to \bar cc) \langle X| O_{c\bar c}  |X\rangle,
\end{eqnarray}
with the $\hat  \sigma(p\bar p\to \bar cc)$ being the partonic cross section. The matrix elements  $
\langle X | O_{\bar cc }  |X\rangle$ are low energy inputs, 
no matter whether  the $X(3872)$ is an ordinary charmonium, a hadron molecule, or a tetraquark. 
Based on the NRQCD framework,  the next-to-leading order calculations~\cite{Butenschoen:2013pxa,Meng:2013gga} are consistent  with the ATLAS data for the production of $X(3872)$ at $\sqrt{s}=8 $TeV~\cite{Aaboud:2016vzw}.

The production mechanism can be further tested in a number of  processes.  For the exclusive 
$e^+e^-\to \gamma X(3872)$ at high energy,   the $\langle X|\bar c \Gamma c|0\rangle$ ($\Gamma$ is a Dirac matrix) contributes, and the cross section should scale as  $1/{s^3}$, derived from the two-quark structure in order to produce the $X(3872)$.  In $B_c\to X(3872)$ decays,   the decay amplitude  is irrespective of the emitted particles in $B_c$ decays. Thus  the ratios of branching fractions of semileptonic and nonleptonic decays, for instance  $B_c\to X\ell\bar\nu$ and $B_c\to  X\rho$,   can be precisely predicted~\cite{Wang:2015rcz} and tested by data.

\subsection{ More on $X(3872)$ production mechanism at hadron level} 
In Ref.~\cite{Bignamini:2009sk}, an inequality for the production rates of $X(3872)$  has been derived: 
\begin{eqnarray}
\sigma(\bar pp\to X)  
&\simeq& \left| \int_{\cal R} d^3{\mathbf k}\langle X|D^0\bar D^{*0}({\mathbf
k}) \rangle\langle D^0\bar D^{*0}({\mathbf k})|\bar pp\rangle\right|^2 \nonumber
\\
&\leq& \int_{\cal R} d^3 {\mathbf k} \left|\Psi({\mathbf k})\right|^2
\int_{\cal R} d^3 {\mathbf k}\left|\langle D^0\bar D^{*0}({\mathbf k})|\bar
pp\rangle\right|^2 \nonumber \\
 &\leq& \int_{\cal R} d^3 {\mathbf k}\left|\langle D^0\bar D^{*0}({\mathbf
k})|\bar pp\rangle\right|^2  
\label{eq:inequality}. 
\end{eqnarray}
References~\cite{Albaladejo:2017blx,Esposito:2017qef} have   discussed the different  choices  of ${\cal R}$ in detail, resulting in dramatically  different conclusions. In the following I will directly  discuss the production mechanism.

The cross section for the inclusive process is defined as  
\begin{eqnarray}
\sigma(\bar pp\to X)   = \int \frac{d^3 p_X}{(2\pi)^32E_X} \frac{d^3 p_{\rm anything}}{(2\pi)^3 2E_{\rm anything}} 
|M_{\bar pp\to X+ anything} |^2 (2\pi)^4\delta^{4}(\sqrt{s}- p_X- p_{\rm anything}),
\end{eqnarray} 
where $X$ denotes the $X(3872)$ and  the symbol ``anything'' denotes the remnant. 

The amplitude is defined  as  (up to some kinematic factor) 
\begin{eqnarray}
M_{\bar pp\to X+ anything}\sim \langle X+ {\rm anything} | T | \bar pp\rangle.
\end{eqnarray}
If one wants to insert a unit operator, one cannot use
\begin{eqnarray}
1=  \int d^3{\mathbf k} |D^0\bar D^{*0}({\mathbf
k}) \rangle\langle D^0\bar D^{*0}({\mathbf k})|,
\end{eqnarray} 
but instead one should use 
\begin{eqnarray}
1=  \int d^3{\mathbf k} d^3{\mathbf p}_{\rm anything '}  |D^0\bar D^{*0}({\mathbf
k}) + {\rm anything}'  \rangle\langle D^0\bar D^{*0}({\mathbf k})+ {\rm anything}' |,
\end{eqnarray} 
where we have   picked up the $D^0\bar D^{*0}({\mathbf
k})$. 
Inserting this unit operator into the matrix element, one has 
 \begin{eqnarray}
M_{\bar pp\to X+ anything}&\sim& \langle X+ {\rm anything} | T | \bar pp\rangle  \nonumber\\
&=& \int d^3{\mathbf k} d^3{\mathbf p}_{\rm anything '}    \langle X+ {\rm anything} | D^0\bar D^{*0}({\mathbf
k}) + {\rm anything}'  \rangle\langle D^0\bar D^{*0}({\mathbf k})+ {\rm anything}' |T | \bar pp\rangle. 
\end{eqnarray}
If one  assumes
\begin{eqnarray}
 {\rm anything} = {\rm anything'},  
\end{eqnarray}  
one will recover the first line of Eq.~\eqref{eq:inequality}. 
However this assumption is not trivial.  An example is the production of $J/\psi$ and other charmonium. If one assumes $ {\rm anything} = {\rm anything'}$, then the $J/\psi$ is only produced by the $\bar cc$ state that has the same quantum numbers as $J/\psi$.  But in the NRQCD approach, it is widely known that the $J/\psi$ can also be  produced by the color octet configurations, in which   $ {\rm anything} \neq {\rm anything'}$. Such contributions are found to be sizable.

The assumption $ {\rm anything} = {\rm anything'}$  for  the production of $X(3872)$ is equivalent to {\it local constituent-molecule duality}, namely, the production rate  of the constituents in the phase space is equivalent to that of the molecule. This is similar to {\it local  quark-hadron duality}, which often fails for very narrow resonances. To recover the quark-hadron duality, one should include final state interactions, which is equivalent to increasing the momentum cutoff~\cite{Artoisenet:2009wk,Artoisenet:2010uu}.

To calculate  the production of ordinary heavy quarkonium,  one often uses  NRQCD, in which a hadron is nonperturbatively produced by  quark fields. Similarly, if there is factorization,  for hadronic molecules we may  establish  an approach  in which the hadron molecule is produced by its constituents, and the low-energy matrix element has been estimated using an effective theory at hadron level. The cross section should have the conjectured  form~\cite{Guo:2014sca,Guo:2014ppa}: 
\begin{eqnarray}
\sigma (p\bar p\to X )&=& \hat  \sigma(p\bar p\to D\bar D^*) \langle X| O_{D \bar D^*}  |X\rangle,
\end{eqnarray}
where $\hat  \sigma(p\bar  p\to D\bar D^*) $ is the {\it partonic} cross section. The  $ \langle X| O_{D \bar D^*}  |X\rangle$  is a  low energy input  and will only be determined in a nonperturbative way.  This approach  avoids the use of {\it local constituent-molecule  duality}, and thus the results should be more reliable.

\section{Conclusion}

The study of production of exotic hadrons is  an important facet of hadron physics. However, in the past decade, there have been great misunderstandings which have hindered us in correctly  understanding the nature of hadron exotics.  
In this work,  
I have demonstrated that 
for a reaction involving hidden flavored hadrons, if there is factorization,   short-distance   and  long-distance degrees of freedom may decouple from each other. Using $e^+e^-\to \rho^0\pi^0$ and a few other examples,  I have shown  that   high energy production does not reveal the hadron's low-energy structure.  This has important consequences in the study of the production of hadron exotics, in particular the $X(3872)$.  This should be a warning to our research community  that the misuse of  production data can   lead to misleading results for the nature of exotic hadrons.

\section*{Acknowledgement} 
I have benefited a lot from   discussions with Profs. Xiangdong Ji, Hsiang-nan Li,  Jian-Ping Ma,  Feng Yuan and Qiang Zhao.  I thank Dr. Jian-Ping Dai for fruitful discussions on $\gamma\gamma \to \pi^+\pi^-$.  I thank the anonymous referee for   critical comments which were valuable in improving the presentation of this work.   This work is supported  in part   by the Thousand Talents Plan for
Young Professionals,  National  Natural
Science Foundation of China under Grant
 No.11575110, 11655002, 11735010, 11747611, Natural  Science Foundation of Shanghai under Grant  No.
15DZ2272100,   and  by   Scientific Research
Foundation for   Returned Overseas Chinese Scholars,   Ministry of Education.
% %%%%%%%%%%%%%%%%%%%%%%%%%%%%%%%%%

\clearpage
\end{CJK*}
\end{document}